\documentclass{article}%
\usepackage[top=3cm, bottom=3cm, left=3cm, right=3cm]{geometry}
\usepackage{amsmath}
\usepackage{amsfonts}
\usepackage{amssymb}
\usepackage{graphicx}
\usepackage{cite}%
\providecommand{\U}[1]{\protect\rule{.1in}{.1in}}
\begin{document}

\title{Multichannel decay: alternative derivation of the $i$-th channel decay probability}
\author{Francesco Giacosa\\\textit{Institute of Physics, Jan Kochanowski University, }\\\textit{ul. Uniwersytecka 7, 25-406, Kielce, Poland}\\\textit{Institute for Theoretical Physics, J. W. Goethe University, }\\\textit{ Max-von-Laue-Str. 1, 60438 Frankfurt, Germany}}
\date{}
\maketitle

\begin{abstract}
In the study of decays, it is quite common that an unstable quantum
state/particle has multiple distinct decay channels. In this case, besides the
survival probability $p(t)$, also the probability $w_{i}(t)$ that the decay
occurs between $(0,t)$ in the $i$-th channel is a relevant object. The general
form of the function $w_{i}(t)$ was recently presented in PLB \textbf{831}
(2022), 137200. Here, we provide a novel and detailed `joint' derivation of
both $p(t)$ and $w_{i}(t)$. As it is well known, $p(t)$ is not an exponential
function; similarly, $w_{i}(t)$ is also not such. In particular, the ratio
$w_{i}/w_{j}$ (for $i\neq j)$ is not a simple constant, as it would be in the
exponential limit. The functions $w_{i}(t)$ and their mutual ratios may
therefore represent a novel tool to study the non-exponential nature of the
decay law.

\end{abstract}

In the study of unstable states, both in QM and in QFT, the survival
probability $p(t)$ (the probability that the state formed at $t=0$ has not
decayed yet at a later time $t>0$) is of crucial importance
\cite{fonda,khalfin,winter,levitan,dicus,peshkin,calderon,koide,fpnew,kk1,duecan,giacosapra1,raczynska,kelkarformal,kelkarlate}%
. Yet, usually unstable states can decay in more than a single decay channel
\cite{pdg}. Then, an equally useful and relevant object is the decay
probability $w_{i}(t)$ that the decay has occurred between $0$ and $t>0$ in a
certain $i$-th channel. Of course, the equality
\begin{equation}
p(t)+\sum_{i=1}^{N}w_{i}(t)=1 \label{trivial}%
\end{equation}
must hold for each $t$, since at any given time the state has either decayed
in one of the $N$ possible channels or it is undecayed (\textit{tertium non
datur}). As it is well established, the survival probability $p(t)$ can be
well approximated by an exponential expression $p(t)\simeq e^{-t/\tau}$ , but
the latter is not exact, as direct and indirect experimental analyses show
\cite{raizen,raizen2,kelkar,rothe,pasclast}. Since $p(t)$ is not an
exponential, it follows that the functions $w_{i}(t)$ are not such as well.

The explicit form for the $w_{i}(t)$ was recently derived in Ref. \cite{mcd}.
A preliminary approximate expression was previously put forward in Ref.
\cite{duecan}. Here, we present a novel joint determination of $p(t)$ and
$w_{i}(t)$ that makes use of a Lippmann-Schwinger equation at the level of
operators, see e.g. Ref. \cite{muether}.

\bigskip

Let $H$ be the Hamiltonian of a physical system that contains an unstable
state $\left\vert S\right\rangle $. We assume that $H$ can be split into
$H=H_{0}+H_{int}$ with $H_{int}=\sum_{i=1}^{N}H_{i},$ where $H_{i}$ is
responsible for the $i$-th decay channel. The ONC eigenstates of the
non-interacting Hamiltonian $H_{0}$ are $\{\left\vert S\right\rangle
,\left\vert E,i\right\rangle \}$: $H_{0}\left\vert S\right\rangle =M\left\vert
S\right\rangle ,$ $H_{0}\left\vert E,i\right\rangle =E\left\vert
E,i\right\rangle $ with $E\geq E_{th,i}$, where $E_{th,i}$ is the energy
threshold of the $i$-th channel; here, we assume for definiteness that
$E_{th,1}\leq E_{th,2}\leq...\leq E_{th,N}$. The ONC conditions of the
underlying Hilbert space read:%
\begin{equation}
\left\langle S|S\right\rangle =1\text{ , }\left\langle S|E,i\right\rangle
=0\text{ , }\left\langle E,i|E^{\prime},j\right\rangle =\delta_{ij}%
\delta(E-E^{\prime})\text{ ;}%
\end{equation}%
\begin{equation}
\left\vert S\right\rangle \left\langle S\right\vert +\sum_{i=1}^{N}%
\int_{E_{th,i}}^{\infty}dE\left\vert E,i\right\rangle \left\langle
E,i\right\vert =1\text{ .} \label{compl}%
\end{equation}
The decays $\left\vert S\right\rangle \rightarrow\left\vert E,i\right\rangle $
are encoded in the matrix elements
\begin{equation}
\left\langle S|H_{j}|E,j\right\rangle =\delta_{ij}\sqrt{\frac{\Gamma_{i}%
(E)}{2\pi}} \label{trans}%
\end{equation}
where $\Gamma_{i}(E)$ is the $i$-th decay width which, in general, is a
function of the energy (it reduces to a constant in the exponential or
Breit-Wigner (BW) limit \cite{ww,ww2,ww3}). [Note, in Eq. (\ref{trans}) a sum
over other d.o.f., such as spin and momenta, has been implicitly taken into
account; the functions $\Gamma_{i}(E)$ are assumed to be known for a specific
quantum system, even though this is usually not a simple task.] An explicit
expression for $H$ that fulfills the properties listed above can be written in
the form of a Friedrichs-Lee Hamiltonian \cite{friedrichs,lee} (for various
applications, see Refs.
\cite{lee2,jc,sherman,gadella,qcdeff,scully,ordonez,xiaozhou1,xiaozhou2,xiaozhou3,leerev,pok,lonigro}
and refs. therein):%
\begin{equation}
H=H_{0}+H_{int},\text{ }%
\end{equation}
with
\begin{equation}
H_{0}=M\left\vert S\right\rangle \left\langle S\right\vert \text{ }+\sum
_{i=1}^{N}\int_{E_{i,th}}^{\infty}\mathrm{dE}E\left\vert E,i\right\rangle
\left\langle E,i\right\vert \text{ , }H_{int}=\sum_{i=1}^{N}\int_{E_{i,th}%
}^{\infty}\mathrm{dE}\sqrt{\frac{\Gamma_{i}(E)}{2\pi}}\left(  \left\vert
E,i\right\rangle \left\langle S\right\vert +\left\vert S\right\rangle
\left\langle E,i\right\vert \right)  \text{.} \label{fl}%
\end{equation}
Note, $H$ represents actually an infinite class of models, since it depends on
the functions $\Gamma_{i}(E).$

The quantity $U(t)=e^{-\frac{i}{\hslash}Ht}$ is the well-known time evolution
operator. In our case, we are interested in the evaluation of the survival
probability amplitude and the $i$-th channel decay amplitude:
\begin{equation}
\left\langle S|U(t)|S\right\rangle \text{ , }\left\langle
E,i|U(t)|S\right\rangle . \label{evt}%
\end{equation}
In order to accomplish it, let us introduce the operator $F(t)$ ($F$ for
`future') as%
\begin{equation}
F(t)=\frac{i}{2\pi}\int_{-\infty}^{+\infty}\mathrm{dE}\frac{e^{-\frac
{i}{\hslash}Et}}{E-H+i\varepsilon}=\left\{
\begin{array}
[c]{c}%
U(t)\text{ for }t>0\\
0\text{ for }t<0
\end{array}
\right.  \text{ .} \label{future}%
\end{equation}
The previous equation should be understood as an an operatorial equation: for
an arbitrary eigenstate $\left\vert E_{0}\right\rangle $ with $H\left\vert
\Psi_{0}\right\rangle =E_{0}\left\vert \Psi_{0}\right\rangle $, one has%
\begin{equation}
F(t)\left\vert \Psi_{0}\right\rangle =\frac{i}{2\pi}\int_{-\infty}^{+\infty
}\mathrm{dE}\frac{e^{-\frac{i}{\hslash}Et}}{E-H+i\varepsilon}\left\vert
\Psi_{0}\right\rangle =\frac{i}{2\pi}\int_{-\infty}^{+\infty}\mathrm{dE}%
\frac{e^{-\frac{i}{\hslash}Et}}{E-E_{0}+i\varepsilon}\left\vert \Psi
_{0}\right\rangle =\left\{
\begin{array}
[c]{c}%
e^{-\frac{i}{\hslash}E_{0}t}\left\vert \Psi_{0}\right\rangle \text{ for }t>0\\
0\text{ for }t<0
\end{array}
\right.  ,
\end{equation}
where the last equation is obtained integrating on the lower half-plane of the
complex variable $E$ for $t>0$ and on the upper half-plane for $t<0.$
Formally, $F(t)$ is not defined for $t=0$ since the integral $\int_{-\infty
}^{+\infty}\mathrm{dE}\frac{1}{E-E_{0}+i\varepsilon}$ does not converge. Yet,
we summarize the previous equation by writing
\begin{equation}
F(t)=\theta(t)U(t)
\end{equation}
together with the choice $\theta(0)=1/2$, thus $F(0)=1/2$. Similarly, let us
introduce the operator $P(t)$ ($P$ for `past'):
\begin{equation}
P(t)=F^{\ast}(-t)=-\frac{i}{2\pi}\int_{-\infty}^{+\infty}\mathrm{dE}%
\frac{e^{-\frac{i}{\hslash}Et}}{E-H-i\varepsilon}=\left\{
\begin{array}
[c]{c}%
0\text{ for }t>0\\
U(t)\text{ for }t<0
\end{array}
\right.  \text{ ,}%
\end{equation}
hence $P(t)=\theta(-t)U(t)$ and $P(0)=1/2$. For each time $t$ (including
$t=0$) we get the consistent result:
\begin{align}
U(t)\text{ }  &  =e^{-\frac{i}{\hslash}Ht}=F(t)+P(t)=\frac{i}{2\pi}%
\int_{-\infty}^{+\infty}\mathrm{dE}\frac{e^{-\frac{i}{\hslash}Et}%
}{E-H+i\varepsilon}-\frac{i}{2\pi}\int_{-\infty}^{+\infty}\mathrm{dE}%
\frac{e^{-\frac{i}{\hslash}Et}}{E-H-i\varepsilon}\nonumber\\
&  =\text{ }\frac{1}{\pi}\int_{-\infty}^{+\infty}\mathrm{dE}\frac{\varepsilon
}{\left(  E-H\right)  ^{2}+\varepsilon^{2}}e^{-\frac{i}{\hslash}Et}%
=\int_{-\infty}^{+\infty}\mathrm{dE}\delta(E-H)e^{-\frac{i}{\hslash}Et}\text{
.}%
\end{align}

\bigskip

Next, we turn to the time evolution of the expectation values of Eq.
(\ref{evt}). In order to evaluate them, we need to determine the propagators
defined as:
\begin{equation}
G_{S}(E)=\left\langle S\right\vert \frac{1}{E-H+i\varepsilon}\left\vert
S\right\rangle \text{ , }T_{i}(E^{\prime},E)=\left\langle E^{\prime
},i\right\vert \frac{1}{E-H+i\varepsilon}\left\vert S\right\rangle \text{ .}%
\end{equation}
Namely, once these quantities are known, the time evolution is obtained by
using the `future' representation $F(t)$ of Eq. (\ref{future}). To this end,
we write down an operatorial Lippmann-Schwinger equation:
\begin{equation}
\frac{1}{E-H+i\varepsilon}=\frac{1}{E-H_{0}+i\varepsilon}+\frac{1}%
{E-H_{0}+i\varepsilon}H_{int}\frac{1}{E-H+i\varepsilon}\text{ ,}%
\end{equation}
which can be proven by considering the operator $O$ defined as (note, dealing
with the operators the ordering is important):%
\begin{align}
O  &  =(E-H_{0}+i\varepsilon)\left(  \frac{1}{E-H+i\varepsilon}-\frac
{1}{E-H_{0}+i\varepsilon}\right)  =(E-H_{0}+i\varepsilon)\frac{1}%
{E-H+i\varepsilon}-1\nonumber\\
&  =(E-H_{0}+i\varepsilon)\frac{1}{E-H+i\varepsilon}-(E-H+i\varepsilon
)\frac{1}{E-H+i\varepsilon}\nonumber\\
&  =(H-H_{0})\frac{1}{E-H+i\varepsilon}=H_{int}\frac{1}{E-H+i\varepsilon
}\text{ .}%
\end{align}
Then, the propagator of the unstable state $S$ reads:
\begin{align}
G_{S}(E)  &  =\left\langle S\right\vert \frac{1}{E-H+i\varepsilon}\left\vert
S\right\rangle =\frac{1}{E-M+i\varepsilon}+\frac{1}{E-M+i\varepsilon
}\left\langle S\right\vert H_{int}\frac{1}{E-H+i\varepsilon}\left\vert
S\right\rangle \nonumber\\
&  =\frac{1}{E-M+i\varepsilon}+\frac{1}{E-M+i\varepsilon}\sum_{i=1}^{N}%
\int_{E_{th,i}}^{\infty}\mathrm{dE}^{\prime}\sqrt{\frac{\Gamma_{i}(E^{\prime
})}{2\pi}}T_{i}(E^{\prime},E)\text{ ,} \label{gs1}%
\end{align}
while the propagators for the transitions $\left\vert S\right\rangle
\rightarrow\left\vert E,i\right\rangle $ are given by:
\begin{equation}
T_{i}(E^{\prime},E)=\left\langle E^{\prime},i\right\vert \frac{1}%
{E-H+i\varepsilon}\left\vert S\right\rangle =\frac{1}{E-E^{\prime
}+i\varepsilon}\left\langle E^{\prime},i\right\vert H_{int}\frac
{1}{E-H+i\varepsilon}\left\vert S\right\rangle =\sqrt{\frac{\Gamma
_{i}(E^{\prime})}{2\pi}}\frac{G_{S}(E)}{E-E^{\prime}+i\varepsilon}\text{ .}
\label{ti1}%
\end{equation}
Plugging $T_{i}(E^{\prime},E)$ into Eq. (\ref{gs1}), we obtain the
Dyson-Schwinger equation of the $S$ propagator:
\begin{equation}
G_{S}(E)=\frac{1}{E-M+i\varepsilon}-\frac{1}{E-M+i\varepsilon}\Pi
(E)G_{S}(E)\text{ ,}%
\end{equation}
where the total self-energy $\Pi(E)$ and the partial self-energies $\Pi
_{i}(E)$ read:
\begin{equation}
\Pi(E)=\sum_{i=1}^{N}\Pi_{i}(E)\text{ , }\Pi_{i}(E)=-\int_{E_{th,i}}^{\infty
}\frac{dE^{\prime}}{2\pi}\frac{\Gamma_{i}(E^{\prime})}{E-E^{\prime
}+i\varepsilon}\text{ ,}%
\end{equation}
for which $\operatorname{Im}\Pi_{i}(E)=\Gamma_{i}(E)/2$ (optical
theorem)\footnote{It is often common to perform the replacements $\Pi
_{i}(E)\rightarrow\Pi_{i}(E)+C_{i},$ where the latter are real subtraction
constants such that $\operatorname{Re}\Pi_{i}(M)=0$. In this way, the bare
mass $M$ of the unstable state is left unchanged by quantum fluctuations.}.
Then:%
\begin{equation}
G_{S}(E)=\frac{1}{E-M+\Pi(E)+i\varepsilon}%
\end{equation}
is the searched propagator of the state $S$. As it is well known, this
expression can be also obtained by performing the standard Dyson resummation,
see e.g. Ref. \cite{leerev}. We thus have provided a simple alternative
derivation of this object.

The propagator $G_{S}(E)$ can be also rewritten as
\begin{equation}
G_{S}(E)=\int_{E_{th,1}}^{+\infty}\mathrm{dE}^{\prime}\frac{d_{S}(E^{\prime}%
)}{E-E^{\prime}+i\varepsilon}\text{ with }d_{S}(E)=-\frac{1}{\pi
}\operatorname{Im}G_{S}(E)=\frac{\Gamma(E)}{2\pi}\left\vert G_{S}%
(E)\right\vert ^{2}\text{ .} \label{spectral}%
\end{equation}
The function $d_{S}(E)$ is the correctly normalized energy distribution (or
spectral function) of the unstable state [$\mathrm{dE}d_{S}(E)$ is the
probability that the state $S$ has an energy between $(E,E+\mathrm{dE})$].
Then, one proceeds as usual for the determination of the survival probability
amplitude:%
\begin{equation}
a(t)=\left\langle S\right\vert U(t)\left\vert S\right\rangle \overset{t>0}%
{=}\left\langle S\right\vert F(t)\left\vert S\right\rangle =\frac{i}{2\pi}%
\int_{-\infty}^{+\infty}\mathrm{dE}G_{S}(E)e^{-\frac{i}{\hslash}Et}%
=\int_{E_{th,1}}^{+\infty}\mathrm{dE}d_{S}(E)e^{-\frac{i}{\hslash}Et}.
\label{ds}%
\end{equation}
This is indeed the amplitude that, starting with $\left\vert S\right\rangle ,$
we still have $\left\vert S\right\rangle $ at the time $t>0.$ The survival
probability%
\begin{equation}
p(t)=\left\vert \int_{E_{th,1}}^{+\infty}\mathrm{dE}d_{S}(E)e^{-\frac
{i}{\hslash}Et}\right\vert ^{2}\text{ } \label{p}%
\end{equation}
emerges. This is indeed the starting point of many studies on the decay law
\cite{fonda,khalfin,winter,levitan,dicus,peshkin,calderon,koide,fpnew,kk1,duecan,giacosapra1,raczynska,kelkarformal,kelkarlate}%
.

As a consequence of the adopted formalism, once $G_{S}(E)$ is fixed, also
$T_{i}(E^{\prime},E)$ in Eq. (\ref{ti1}) is determined. We then calculate the
probability that the decay takes place in the $i$-th channel between $0$ and
$t>0$ as:%
\begin{align}
w_{i}(t)  &  =\int_{E_{th,1}}^{+\infty}dE^{\prime}\left\vert \left\langle
E^{\prime},i\right\vert U(t)\left\vert S\right\rangle \right\vert ^{2}%
\overset{t>0}{=}\int_{E_{th,1}}^{+\infty}dE^{\prime}\left\vert \left\langle
E^{\prime},i\right\vert F(t)\left\vert S\right\rangle \right\vert
^{2}\nonumber\\
&  =\int_{E_{th,1}}^{+\infty}dE^{\prime}\left\vert \frac{i}{2\pi}\int
_{-\infty}^{+\infty}dET_{i}(E^{\prime},E)e^{-\frac{i}{\hslash}Et}\right\vert
^{2}=\int_{E_{th,1}}^{+\infty}dE^{\prime}\frac{\Gamma_{i}(E^{\prime})}{2\pi
}\left\vert \frac{i}{2\pi}\int_{-\infty}^{+\infty}dE\frac{G_{S}(E)}%
{E-E^{\prime}+i\varepsilon}e^{-\frac{i}{\hslash}Et}\right\vert ^{2}\text{ .}
\label{wi}%
\end{align}
This is indeed the expression for the quantity $w_{i}(t)$ that we were looking
for. Yet, it still involves the complex propagator $G_{S}(E)$, so it is better
to recast it in a form that is simpler in practical applications. By
introducing the spectral representation of Eq. (\ref{spectral})
\begin{align}
\frac{i}{2\pi}\int_{-\infty}^{+\infty}\frac{G_{S}(E)}{E-E^{\prime
}+i\varepsilon}e^{-\frac{i}{\hslash}Et}  &  =\frac{i}{2\pi}\int_{-\infty
}^{+\infty}dE\int_{E_{1,th}}^{+\infty}\mathrm{dy}\frac{1}{E-E^{\prime
}+i\varepsilon}\frac{d_{S}(y)}{E-y+i\varepsilon}e^{-\frac{i}{\hslash}%
Et}\nonumber\\
&  =\int_{E_{1h,1}}^{+\infty}\mathrm{dy}d_{S}(y)\left(  \frac{e^{-\frac
{i}{\hslash}E^{\prime}t}-e^{-\frac{i}{\hslash}yt}}{E^{\prime}-y}\right)
\end{align}
(note, the integrand contains no singularity) we obtain the expression of Ref.
\cite{mcd}:
\begin{equation}
w_{i}(t)=\int_{E_{th,1}}^{+\infty}dE^{\prime}\frac{\Gamma_{i}(E^{\prime}%
)}{2\pi}\left\vert \int_{E_{1,th}}^{+\infty}\mathrm{dy}d_{S}(y)\left(
\frac{e^{-\frac{i}{\hslash}E^{\prime}t}-e^{-\frac{i}{\hslash}y^{\prime}t}%
}{E^{\prime}-y}\right)  \right\vert ^{2}\text{ .}%
\end{equation}
This quantity can be calculated numerically when the functions $\Gamma_{i}(E)$
(and thus $d_{S}(E)$ too) are known. Roughly speaking, it is ready to be used
by just \textquotedblleft plug in and calculate\textquotedblright.

There is another useful way to express $w_{i}(t)$ mentioned in Ref.
\cite{mcd}. By introducing
\begin{align}
I(t)  &  =\int_{0}^{t}\mathrm{dt}^{\prime}\frac{a(t^{\prime})}{\hslash
}e^{\frac{i}{\hslash}E^{\prime}t^{\prime}}=\int_{0}^{t}dt^{\prime}\left(
\int_{E_{th,1}}^{+\infty}\mathrm{dy}d_{S}(y)e^{-\frac{i}{\hslash}yt^{\prime}%
}\right)  e^{\frac{i}{\hslash}E^{\prime}t^{\prime}}=\int_{E_{th,1}}^{+\infty
}\frac{\mathrm{dy}}{\hslash}d_{S}(y)\int_{0}^{t}dt^{\prime}e^{\frac{i}%
{\hslash}(E^{\prime}-y)t^{\prime}}\nonumber\\
&  =\int_{-\infty}^{+\infty}\mathrm{dy}d_{S}(y)\frac{e^{\frac{i}{\hslash
}(E^{\prime}-y)t}-1}{i(E^{\prime}-y)}=ie^{\frac{i}{\hslash}E^{\prime}t}%
\int_{E_{th,1}}^{+\infty}\mathrm{dy}d_{S}(y)\frac{e^{-\frac{i}{\hslash
}E^{\prime}t}-e^{\frac{i}{\hslash}y}}{E^{\prime}-y}%
\end{align}
we find%
\begin{equation}
w_{i}(t)=\int_{E_{th,i}}^{+\infty}\mathrm{dE}^{\prime}\frac{\Gamma
_{i}(E^{\prime})}{2\pi}\left\vert \int_{0}^{t}dt^{\prime}\frac{a(t^{\prime}%
)}{\hslash}e^{\frac{i}{\hslash}E^{\prime}t^{\prime}}\right\vert ^{2}\text{ }.
\end{equation}
Once $a(t)$ is calculated (a necessary step for getting the survival
probability $p(t)$), $w_{i}(t)$ can be numerically evaluated from the previous expression.

Next, we recall some relevant properties and extensions:

\begin{itemize}
\item We can prove Eq. (\ref{trivial}) by using the formal expression for the
transitions $w_{i}(t)$ in Eq. (\ref{wi}) and the completeness relation of Eq.
(\ref{compl}):
\end{itemize}

\begin{align}
\sum_{i=1}^{N}w_{i}(t)  &  =\sum_{i=1}^{N}\int_{E_{th,i}}^{+\infty}%
\mathrm{dE}^{\prime}\left\vert \left\langle E^{\prime},i\right\vert
U(t)\left\vert S\right\rangle \right\vert ^{2}=\left\langle S\right\vert
U^{\dag}(t)\left(  \sum_{i=1}^{N}\int_{E_{th,i}}^{+\infty}\mathrm{dE}^{\prime
}\left\vert E^{\prime},i\right\rangle \left\langle E^{\prime},i\right\vert
\right)  U(t)\left\vert S\right\rangle \nonumber\\
&  =\left\langle S\right\vert U^{\dag}(t)\left(  1-\left\vert S\right\rangle
\left\langle S\right\vert \right)  U(t)\left\vert S\right\rangle =1-p(t)\text{
.}%
\end{align}

This is an important consistency check of the correctness of the obtained results.

\begin{itemize}
\item The exponential (or Breit-Wigner) limit \cite{ww,ww2,ww3} is obtained
for $\Gamma_{i}=const$ and $\Gamma=\sum_{i=1}^{N}\Gamma_{i}$ (no energy
dependence). The survival probability $p(t)$ and the decay probabilities
$w_{i}(t)$ reduce to \cite{duecan,mcd}:
\begin{equation}
p(t)=e^{-\frac{\Gamma}{\hslash}t}\text{ , }w_{i}(t)=\frac{\Gamma_{i}}{\Gamma
}\left(  1-e^{-\frac{\Gamma}{\hslash}t}\right)  \text{ }\rightarrow\frac
{w_{i}(t)}{w_{j}(t)}=\frac{\Gamma_{i}}{\Gamma_{j}}=const\text{ .} \label{bw2}%
\end{equation}

\item In the general case, the ratio $w_{i}/w_{j}\neq const$ (for $i\neq j$).
This fact has been shown in\ Ref. \cite{mcd} by using the widths $\Gamma
_{i}(E)=2g_{i}^{2}\frac{\sqrt{E-E_{th,i}}}{E^{2}+\Lambda^{2}}$ inspired by the
expressions derived in Ref. \cite{pascazioidrogeno} in the case of
hydrogen-like atoms. In Ref. \cite{duecan} $w_{i}/w_{j}$ was also shown to be
not a simple constant (in the framework of an approximate solution) for
various choices of $\Gamma_{i}(E)$.

\item A related interesting quantity is $h_{i}(t)=w_{i}^{\prime}(t)$:
$h_{i}(t)dt$ is the probability that the decay takes place in the $i$-th
channel in the interval $(t,t+dt).$ In the BW limit, $h_{i}(t)/h_{j}%
(t)=\Gamma_{i}/\Gamma_{j}=const,$ but this does not apply in general
\cite{duecan,mcd}.

\item In Ref. \cite{2delta} the two-channel decay was studied by in the
framework of an asymmetric double-delta potential $V(x)=V_{0}\left(
\delta(x-a\right)  +k\delta(x+a),$ where $k\neq1$: the two channels were
represented by the tunneling to the `left' and to the `right'. The numerical
accurate solutions of the Schr\"{o}dinger equation has clearly shown that
$w_{R}(t)/w_{L}(t)$ as well as $h_{R}(t)/h_{L}(t)$ (where $R$ stays for right
and $L$ left) are not constant.

\item The results can be extended to QFT. For that, the variable $E$ must be
replaced with $s=E^{2}$ (for the relativistic version of the Friedrichs-Lee
approach, see e.g. Refs. \cite{prigoginerel,zhourel1,zhourel2} ). The
propagator reads $G_{S}(s)=\left[  s-M^{2}+\Pi(s)+i\varepsilon\right]  ^{-1}$,
where $\Pi(s)=\sum_{i=1}^{N}\Pi_{i}(s)$ (with $\operatorname{Im}\Pi
_{i}(s)=\sqrt{s}\Gamma_{i}(s))$ is the sum of the self energies for $N$
distinct decay channels. The spectral function is $d_{S}(s)=-\frac{1}{\pi
}\operatorname{Im}G_{S}(s)$ (e.g. Refs. \cite{salam,lupo}). The survival
probability $p(t)$ takes an analogous form of Eq. (\ref{p}) (e.g. Refs.
\cite{nonexpqft,vanhove}):%
\begin{equation}
p^{\text{QFT}}(t)=\left\vert \int_{s_{th,1}}^{+\infty}\mathrm{ds}%
d_{S}(s)e^{-\frac{i}{\hslash}\sqrt{s}t}\right\vert ^{2}\text{ ,}%
\end{equation}
while the partial decay probability $w_{i}(t)$ read:
\end{itemize}

\begin{equation}
w_{i}^{\text{QFT}}(t)=\int_{s_{th,i}}^{+\infty}\mathrm{ds}\frac{\sqrt{s}%
\Gamma_{i}(s)}{\pi}\left\vert \int_{s_{th,1}}^{+\infty}\mathrm{ds}^{\prime
}d_{S}(s^{\prime})\left(  \frac{e^{-\frac{i}{\hslash}\sqrt{s}t}-e^{-\frac
{i}{\hslash}\sqrt{s^{\prime}}t}}{s-s^{\prime}}\right)  \right\vert ^{2}\text{
.}%
\end{equation}

This expression can be calculated numerically once the functions $\Gamma
_{i}(s)$ are known.

\begin{itemize}
\item In QFT, there is no BW limit and no exponential decay (a threshold is
always present, since $s\geq0$). Setting $\Gamma_{i}(s)$ to a constant leads
to some inconsistencies. An interesting model, discussed in Ref. \cite{sill},
postulates $\Pi_{i}(s)=i\tilde{\Gamma}_{i}\sqrt{s-s_{th,i}},$ for which
$\Gamma_{i}(s)=\tilde{\Gamma}_{i}\sqrt{(s-s_{th,i})/s}\theta\left(
s-s_{th,i}\right)  $ (that reduces to a constant for large $s$). Despite its
simplicity, it allows to fit quite well the spectral functions of various
broad hadrons. The function $w_{i}(t)$ turns out to be, as expected,
non-exponential, in agreement wit the QM case.
\end{itemize}

\bigskip

In conclusion, we have presented a novel and simple way to obtain the
expressions of the survival probability $p(t)$ and the decay probability in
the $i$-th channel $w_{i}(t)$ by using a Lippmann-Schwinger equation at the
level of operators. The propagator for the state $S$ and the transition
propagator for $S$ into any decay product are intertwined. In this way, $p(t)$
and $w_{i}(t)$ naturally emerge and the results coincide with the ones shown
in\ Ref. \cite{mcd}. In the future, the study of $w_{i}(t)$ in various
physical systems is planned.

\bigskip

\textbf{Acknowledgments:} the author thanks L. Tinti, G. Pagliara and S.
Mr\'{o}wczy\'{n}ski for stimulating and useful discussions. Financial support
from the OPUS project 2019/33/B/ST2/00613 is acknowledged.

\end{document}